\documentclass[copyright,creativecommons]{eptcs}
\providecommand{\event}{ACL2 Workshop 2015 (ACL2`15)} 
\usepackage{breakurl}             

\usepackage{caption}
\usepackage{epsfig}
\usepackage{float}
\usepackage{graphicx}
\usepackage{amsmath}
\usepackage{amsfonts}
\usepackage{amssymb}
\usepackage{xspace}
\usepackage{hyperref}
\newcommand{\smtlink}{\texttt{Smtlink}\xspace}
\newcommand{\cnfsmtlink}{\texttt{Smtlink}\xspace}
\newcommand{\cnfsmtlinkc}{\texttt{Smtlink-custom-config}\xspace}
\newcommand{\gsmt}{\ensuremath{G_{\mathit{SMT}}}}
\newcommand{\hide}[1]{}

\usepackage{listings}
\usepackage{color}

\definecolor{dkgreen}{rgb}{0,0.6,0}
\definecolor{gray}{rgb}{0.5,0.5,0.5}
\definecolor{mauve}{rgb}{0.58,0,0.82}

\graphicspath{{./figs/}}
 
\lstdefinestyle{mystyle}{
    aboveskip=3mm,
    belowskip=3mm,
    showstringspaces=false,
    columns=flexible,
    basicstyle={\small\ttfamily},
    numbers=left,
    numbers=left,
    numbersep=5pt,
    numberstyle=\tiny\color{gray},
    keywordstyle=\color{blue},
    commentstyle=\color{dkgreen},
    stringstyle=\color{mauve},
    breaklines=true,
    breakatwhitespace=true,
    tabsize=3,
    keepspaces=true
}
 
\title{Extending ACL2 with SMT Solvers}
\author{
Yan Peng \qquad\qquad Mark Greenstreet
\institute{University of British Columbia\\
Vancouver, Canada}
\email{\quad {yanpeng,mrg}@cs.ubc.ca}
}
\def\titlerunning{Extending ACL2 with SMT Solvers}
\def\authorrunning{Y. Peng \& M. Greenstreet}

\floatstyle{ruled}
\newfloat{Program}{htbp}{lop}[section]

\begin{document}
\maketitle

\begin{abstract}
We present our extension of ACL2 with Satisfiability Modulo Theories (SMT) solvers using ACL2's trusted clause processor mechanism. We are particularly interested in the verification of physical systems including Analog and Mixed-Signal (AMS) designs. ACL2 offers strong induction abilities for reasoning about sequences and SMT complements deduction methods like ACL2 with fast nonlinear arithmetic solving procedures. While SAT solvers have been integrated into ACL2 in previous work, SMT methods raise new issues because of their support for a broader range of domains including real numbers and uninterpreted functions. This paper presents \smtlink, our clause processor for integrating SMT solvers into ACL2. We describe key design and implementation issues and describe our experience with its use. 
\end{abstract}

\section{Introduction}
This paper presents \smtlink{}, a clause processor for using satisfiability modulo theory (SMT)
solvers to discharge proof goals in ACL2.
Prior work has~\cite{Reeber06,Swords11} incorporated SAT solving into ACL2,
and Manolios and Srinivasan~\cite{Manolios2006,srinivasan2007} described an extension of ACL2 with the Yices SMT solver.
Our work explores the use of SMT solvers for their decision procedures for linear and non-linear arithmetic
which, to the best of our knowledge, has not been addressed in prior work.

Interactive theorem proving and SMT solving provide complementary strengths for verification.
SMT solvers can automatically discharge proof obligations that would be tedious to handle with an
interactive theorem prover alone.
Conversely, theorem provers provide methods for proof by induction and proof structuring methods.
While there has been some work on automatically proving induction proofs using SMT solvers
(see~\cite{Rustan12}), theorem provers such as ACL2 offer a much more comprehensive framework for induction
proofs.
For many problems, SMT solvers cannot prove the main result in a single step; in fact, the main
theorem may not even be expressible in the logic of the SMT solver.  However, the SMT solver can
discharge key lemmas to simplify the proof process, and the theorem prover can ensure that the
proofs for the main theorems are, indeed, complete.
When used from within an interactive theorem prover, the user can identify key goals and \emph{relevant}
facts to make effective use of the SMT solver.
Doing so can avoid sending the SMT solver down a path of an intractable number of useless branches
and lead instead to a proof of the desired goal.

Our intended application of the combination of ACL2 with an SMT solver is to verify properties
of Analog and Mixed-Signal (AMS) circuits and other cyber-physical systems.
AMS circuits are mixed analog and digital systems, typically consisting of multiple analog
and digital feedback loops operating at much different time scales.
It is not practical to simulate AMS circuits for all possible device parameters,
initial conditions, inputs, and operating conditions. In fact, running
just one such simulation may require more time than the design schedule.
Most AMS circuits are intended to be correct for relatively simple reasons
–- errors occur because the designer's informal reasoning overlooked some
critical case or had some simple error. Our approach is to verify that the
intuitive argument for correctness is indeed correct by reproducing the argument
in an automated, interactive theorem prover, ACL2. The advantage of using a
theorem prover is soundness and generality: by using a carefully designed
and thoroughly tested theorem prover, we have high confidence in the theorems that it establishes. The critical limitation of using a theorem prover is
that formulating the proofs can require large amounts of very highly skilled
effort. Our solution is to integrate a SMT solver, Z3, into ACL2. This allows many parts of the
proof, especially those involving large amounts of tedious algebra, to be
performed automatically.
While our focus is on AMS, the issues addressed here are common to those in most computing devices and other physical systems.

Our implementation uses ACL2's trusted clause processor mechanism for integrating external procedures.
Our goal is to provide a flexible framework for developing proofs in a relatively new application domain.
Thus, our clause processor is designed to be easily configured and modified by the user.
However, too much freedom to change the behaviour of the clause processor also raises the spectre of
unsoundness.  We address this with a two-pronged solution.
Our clause processor is available with a standard configuration, where the soundness depends mainly on the
soundness of ACL2, the SMT solver, and a small amount of interface code.
There is also a customizable configuration that has a separate trust-tag.
This facilitates experimentation, but places the burden for soundness directly upon the user.
We describe our use of the two approaches, and show how this combination provides a flexible
environment for experimentation and a safe environment for ``production'' use.

The key contributions of this work are:
\begin{itemize}
  \item We present our software architecture for integrating an SMT solver into ACL2 as a trusted clause
  	  processor.
  \item We describe the issues that arose in this integration, our solutions, and the rationale behind
          our design choices.
  \item Our emphasis is on using the arithmetic capabilities of the Z3 SMT solver.  This differs from
          most prior work on integrating SMT solvers into theorem provers that has focused on using
	  decision procedures for SAT, integer arithmetic, and discrete data structures.
  \item We show how some simple customizations of the general framework can lead to a dramatic
          reduction in proof effort.
\end{itemize}

The rest of this paper is organized as follows: Section~\ref{sec:tour} introduces our clause processor with
three simple examples.
Section~\ref{sec:arch} describes our software architecture, the issues that arise when integrating an SMT
solver into ACL2, and our solutions to these issues.
Section~\ref{sec:custom} describes how the SMT interface can be customized.  In particular, we show how
adding a simple inference engine that provides an incomplete theory of \texttt{expt} greatly simplifies
our proofs for verifying properties of an AMS circuit.
Sections~\ref{sec:relwk} and~\ref{sec:concl} present related work and a summary of the current work
respectively.

\section{A Short Tour}\label{sec:tour}
This section presents simple theorems that can be proven using \smtlink.
The examples here assume that the \smtlink{} book has been downloaded from:\\
\rule{2em}{0ex}\url{https://bitbucket.org/pennyan/smtlink}\\
and certified using \texttt{cert.pl} (see the instructions in the \texttt{README} file).
Program~\ref{prog:smtlink-include} shows how to include the \smtlink{} book
\begin{Program}[t]
  \caption{Including the \smtlink{} book}
  \label{prog:smtlink-include}
\begin{lstlisting}[language=Lisp]
  (add-include-book-dir :cp "/dir/to/smtlink")
  (include-book "top" :dir :cp)
  (tshell-ensure)
\end{lstlisting}
\end{Program}
where \texttt{/dir/to/smtlink} is the directory with the \smtlink{} book.
The \texttt{(tshell-ensure)} form allows \smtlink{} to invoke the SMT solver in a separate process.
\smtlink supports two configurations.
The examples in this section use \cnfsmtlink, which uses default settings.
The other, \cnfsmtlinkc, can be configured by the user and is described in Section~\ref{sec:custom}.

\begin{Program}[t]
  \caption{A theorem about a system of polynomial inequalities}
  \label{prog:poly-ineq}
\begin{lstlisting}[language=Lisp]
(defthm poly-ineq-example-a
  (implies (and (rationalp x) (rationalp y)
                (<= (+ (* 4/5 x x) (* y y)) 1)
                (<= (- (* x x) (* y y)) 1))
           (<= y (- (* 3 (- x 17/8) (- x 17/8)) 3)))
  :hints (("Goal"
           :clause-processor
           (Smtlink clause nil))))

(defthm poly-ineq-example-b
  (implies (and (rationalp x) (rationalp y)
                (<= (+ (* 2/3 x x) (* y y)) 1)
                (<= (- (* x x) (* y y)) 1))
           (<= y (+ 2
                    (- (* 4/9 x))
                    (- (* x x x x))
                    (* 1/4 x x x x x x)) ))
  :hints (("Goal"
           :clause-processor
           (Smtlink clause nil))))
\end{lstlisting}
\end{Program}
Program~\ref{prog:poly-ineq} shows two examples involving systems of polynomial inequalities:
\texttt{nil} is a list of additional hints for the clause processor as no further hints are needed for these examples.
Why would we want to prove such theorems?
Simple, they illustrate the challenges of using ACL2 to reason about systems of polynomial inequalities
as often appear in models of physical systems including AMS verification.
Without the clause-processor, the proofs fail in ACL2 with the \texttt{:nonlinearp} hint enabled
and with or without any of the arithmetic books (i.e.\ \texttt{arithmetic/top-with-meta},
\texttt{arithmetic-2/meta/top}, \texttt{arithmetic-3/top}, and or \texttt{arithmetic5/top}).
Of course, a patient and savvy user could guide ACL2 through a sequence of lemmas and eventually discharge the claims.
Using the SMT solver, the theorems are proven automatically.

Some theorems, while tedious to prove in ACL2, simply cannot be proven by SMT techniques alone.
Consider Program~\ref{prog:poly-of-expt}.
\begin{Program}[t]
  \caption{A claim with non-polynomial arithmetic}\label{prog:poly-of-expt}
\begin{lstlisting}[language=Lisp]
(defthm poly-of-expt-example
  (implies (and (rationalp x) (rationalp y) (rationalp z) (integerp m)
                (integerp n) (< 0 z) (< z 1) (< 0 m) (< m n))
           (<= (* 2 (expt z n) x y) (* (expt z m) (+ (* x x) (* y y)))))
  :hints(("Goal"
          :clause-processor
          (Smtlink clause '((:let ((expt_z_m (expt z m) rationalp)
                                   (expt_z_n (expt z n) rationalp)))
                            (:hypothesize ((< expt_z_n expt_z_m)
                                           (< 0 expt_z_m)
                                           (< 0 expt_z_n)))
                  ) ))))
\end{lstlisting}
\end{Program}
Again, when just using ACL2, 
the proof fails with or without a \texttt{:nonlinearp} hint or any of the arithmetic books.
As formulated, \texttt{poly-of-expt-example} would appear to be unsuitable for proof with our SMT techniques
because we are using Z3 as our SMT solver, and Z3 does not support reasoning about non-polynomial
functions such as \texttt{expt}.
Our solution is to allow the user to give hints to the clause processor.
These hints allow the user to direct the clause processor to replace all occurrences of a given expression with a new, free variable,
and to express constraints that are satisfied by these variables.
A complete description of these hints is presented in Section~\ref{sec:arch}.
To prove \texttt{poly-of-expt-example}, we use the clause-processor hint.
We also include the book \texttt{arithmetic-5/top}.
The two \texttt{:let} hints direct \smtlink{} to replace all occurrences of \texttt{(expt z m)} with the variable \texttt{expt\_z\_m};
furthermore, we are asserting that the value \texttt{(expt z m)} satisfies \texttt{rationalp}.
Likewise for {\texttt{expt\_z\_n}} replacing all occurrences of \texttt{(expt z n)}.
The three \texttt{:hypothesize} hints state additional constraints on the values of \texttt{expt\_z\_m} and \texttt{expt\_z\_m} for use by the SMT solver.
With these substitutions and constraints, Z3 readily discharges the main claim.

For this approach to be sound, these substitions, type-assertions, and constraints must all be
implied by the hypotheses of the original theorem.
If the SMT solver discharges the main claim, then \smtlink{} returns each of these added assumptions and new clauses to be proven by ACL2.
In other words, \smtlink{} has replaced a clause that would be difficult to prove using ACL2 alone,
with a moderate number of simpler clauses that are simpler for ACL2 to establish,
plus one clause (the augmented, original claim) that is proven by the SMT solver. 
In this case, runes from \texttt{arithmetic-5/top} enable the returned clauses to be discharged without further assistance.
This also illustrates the synergies that are available by combining SMT techniques with theorem proving.

\begin{figure}[t]
  \begin{center}
    \resizebox{0.9\textwidth}{!}{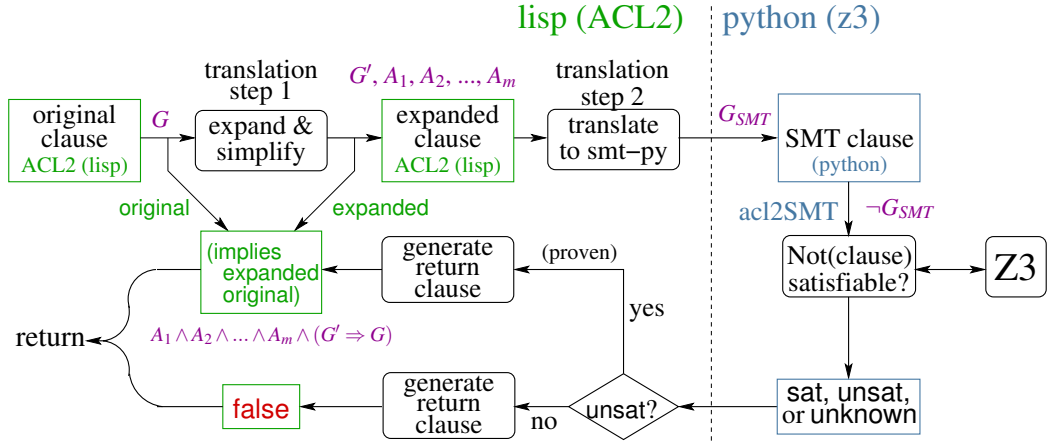}
    \caption[Clause processor top architecture]{\label{fig:cp-top} Top-level architecture of \smtlink}
  \end{center}
\end{figure}
\section{Software architecture of \smtlink}\label{sec:arch}
Figure~\ref{fig:cp-top} shows the structure of \smtlink{}.
The clause processor translates ACL2 clauses into a Python representation inspired by Z3's Python API.
The translation process is divided into two phases.
The first phase translates from ACL2 to ACL2.
This translation allows the clause-processor to accept a fairly expressive subset of the ACL2 language
while the expanded clauses output by this phase use only a small set of primitive Lisp functions
(See Section~\ref{sec:arch.phase2.functions}).
The second phase translates the simplified (but expanded) ACL2 clauses to our Python API -- this process
is the main ``trusted'' aspect of our trusted clause processor.
The SMT solver verifies a clause by showing that its negation is unsatisfiable.
If this is the case, then \smtlink{} returns a list of clauses for subgoals that arose
in the translation process.
Essentially, \smtlink{} asks ACL2 to verify that the expanded clause implies the original,
and to verify any type assertions or additional hypotheses that were provided by the user.
If the SMT solver fails to show that the clause is unsatisfiable,
it typically provides a counter-example that \smtlink{} then prints to the ACL2 comment window,
although is some cases it may simply report that the satisfiability of the clause is ``unknown''.
In these cases, \smtlink{} prints the counter-example or ``unknown'' status to the ACL2 comment window
and aborts the proof attempt.

\subsection{The first translation phase}\label{sec:arch.phase1}
The first phase of translation transforms clauses written in a fairly expressive subset of ACL2
into a very small subset.
Most of the complexity of the translation process is in this first phase.
As described in Section~\ref{sec:arch.sound}, \smtlink{} constructs a new clause that is proven by
ACL2 to validate this translation.
The key issues in the first phase are:
\begin{itemize}
  \item ACL2 is untyped whereas SMT solvers support many-sorted logics.
  \item ACL2 clauses often include user-defined functions.
  \item The user may add type assertions and/or extra hypotheses to enable the SMT solver to discharge
  	  a claim.  These must be verified by ACL2.
  \item The user may need to provide hints to enable ACL2 to discharge subgoals that are returned by
    the clause processor.
\end{itemize}

\begin{Program}[t]
  \caption{A putative theorem without type constraints}\label{prog:not-really-a-theorem}
\begin{lstlisting}[language=Lisp]
(defthm not-really-a-theorem
  (iff (equal x y) (zerop (- x y))) )
\end{lstlisting}
\end{Program}
\begin{Program}[t]
  \caption{A simple theorem with type constraints}\label{prog:rational-minus-and-equal}
\begin{lstlisting}[language=Lisp]
(defthm rational-minus-and-equal
  (implies (and (rationalp x) (rationalp y))
           (iff (equal x y) (zerop (- x y))) ))
\end{lstlisting}
\end{Program}
\subsubsection{Types}\label{sec:arch.phase1.types}
Consider the putative theorem shown in Program~\ref{prog:not-really-a-theorem}.
ACL2 is untyped and requires all functions to be total.
Accordingly, \texttt{(- x y)} is defined for all values for \texttt{x} and \texttt{y},
including non-numeric values.
As defined in ACL2, arithmetic operators such as \texttt{-} treat non-numeric values as if they were \texttt{0}.
Thus, \texttt{x $=$ 'dog} and \texttt{y $=$ (list "hello" 2 'world)} is a counter-example to
\texttt{not-really-a-theorem}.
On the other hand, Z3 uses a typed logic, and each variable must have an associated sort.
If we treat \texttt{x} and \texttt{y} as real-valued variables,
the \texttt{z3py} equivalent to \texttt{not-really-a-theorem} is
{\tt\begin{tabbing}xx\=\+\kill
  >>> x, y = Reals(['x', 'y'])\\
  >>> prove((x == y) == ((x - y) == 0))\\
  {\color{blue} proved}
\end{tabbing}}\noindent
In other words, \texttt{not-really-a-theorem} as expressed in the untyped logic of ACL2 is not a theorem,
but the ``best'' approximation we can make in the many-sorted logic of Z3 is a theorem.
To solve these problems, \smtlink{} requires that each free variable in a theorem is constrained
by an ACL2 type recognizer such as \texttt{integerp} and \texttt{rationalp}.
These are then translated to corresponding SMT sorts with the design requirement that the set of
values in the SMT sort must be a superset (or equal to) the set of values admitted by the type
recognizer.

Although ACL2 is untyped, it is common for users to include assertions such as \texttt{(rationalp x)}
that constrain the types of free-variables appearing in a theorem.
Program~\ref{prog:rational-minus-and-equal} shows the previous, putative theorem
with type recognizers added to the hypotheses.
ACL2 proves \texttt{rational-minus-and-equal} without any assistance from the user.
Note that \texttt{rational-minus-and-equal} holds for \emph{all} values of \texttt{x} and \texttt{y}
including values that are not rational, and values that are not even numeric, such as
\texttt{x $=$ 'dog} and \texttt{y $=$ (list "hello" 2 'world)}.
For such cases, the antecedent of the theorem is not satisfied, and the theorem holds vacuously.

Let $G$ be the clause to be proven by \smtlink{}; $G$ is the ``goal''.
In the first translation phase, \smtlink{} traverses $G$ looking for terms of the form
\textit{(typep var)} where \textit{typep} is one of \texttt{booleanp}, \texttt{integerp}, or 
\texttt{rationalp}; \textit{var} is a symbol (but not nil); and the clause holds vacuously if
\texttt{(not\textit{~(typep var)})}.  In other words, such terms are \emph{type hypotheses}.
\smtlink{} identifies type hypotheses \textit{syntactically} by walking the tree for the
expression, recognizing the constructions for \texttt{if}, \texttt{implies}, \texttt{not},
and the type-recognizers (note: the ACL2 macros ``\texttt{and}'' and ``\texttt{or}'' expand to terms
written with \texttt{if}).

\hide{
Let $T = \texttt{(list $T_1$ $T_2$ \ldots $T_m$)}$ be the list of all type-hypotheses;
$\widehat{T}$ denote the conjunction of the elements of $T$; and
and $G_T$ be $G$ rewritten by replacing each of the $T_i$'s
with the boolean constant \texttt{t}.
\smtlink{} constructs the terms $\widehat{T} \Rightarrow G_T$, $\widehat{T} \vee G$, and
$((\widehat{T} \vee G) \wedge (\widehat{T} \Rightarrow G_T)) \Rightarrow G$.  
\smtlink{} invokes the SMT solver to determine if $G_T$ holds for all valuations of the free variables
that satisfy $T$.  If the SMT solver can show this, then $\widehat{T} \Rightarrow G_T$ is established.
Then, the terms $\widehat{T} \vee G$, and $((\widehat{T} \vee G) \wedge (\widehat{T} \Rightarrow G_T)) \Rightarrow G$
are returned to ACL2 to be proven.
Observe that this proves $G$ for \emph{any} choice of $\widehat{T}$ and $G_T$.
This is how we make the soundness of our approach independent of the operations in the first translation phase.
Of course, getting the \emph{intended} $T$ and $G_T$ are crucial for proving the subgoals.
The remainder of this section describes the other operations performed in the first translation phase,
and Section~\ref{sec:arch.sound} shows how the approach described in this paragraph for using ACL2 to check
\smtlink{}'s derivation of type hypotheses is extended to the entire translation process.
}

Let $T = \texttt{(list $T_1$ $T_2$ \ldots $T_m$)}$ be the list of all type-hypotheses;
$\widehat{T}$ denote the conjunction of the elements of $T$; and
and $G_T$ be $G$ rewritten by replacing each of the $T_i$'s
with the boolean constant \texttt{t}.
We could now construct the terms $\widehat{T} \Rightarrow G_T$, $\widehat{T} \vee G$, and
$((\widehat{T} \vee G) \wedge (\widehat{T} \Rightarrow G_T)) \Rightarrow G$.
We could then invoke the SMT solver to determine if $G_T$ holds for all valuations of the free variables
that satisfy $T$.
If the SMT solver can show this, then $\widehat{T} \Rightarrow G_T$ is established.
Then, we could return the terms $\widehat{T} \vee G$, and $((\widehat{T} \vee G) \wedge (\widehat{T} \Rightarrow G_T)) \Rightarrow G$
to ACL2 to be proven.
If these proofs are successful, then we can conclude that $G$ is a theorem as well.
\smtlink{} uses this approach; however, rather than checking each step of the first translation phase,
it checks the final result.
Section~\ref{sec:arch.sound} describes this process.
		  
\subsubsection{Functions}\label{sec:arch.phase1.functions}
The second phase of translation supports a small set of ACL2 built-in functions
(see Section \ref{sec:arch.phase2}).
\smtlink{} handles other functions by expanding their calls.
In particular \textit{(fun actual-parameters)} becomes
\begin{equation}\label{eq:fun-expand}
  \mathtt{((lambda\;(\textit{fresh-variables-for-formals})\;\textit{body-of-fun})\ \textit{actual-parameters})}
\end{equation}
Because \textit{body-of-fun} may have function instances that need to be expanded,
\smtlink{} recursively applies this function-expansion operation to \textit{body-of-fun} and each term in \textit{actual-parameters}.

If the function \textit{fun} has a recursive definition, then the expansion procedure described will not terminate.
To avoid this problem, we require the user to specify a maximum expansion depth and the return type
for each function.
\smtlink{} replaces each call beyond the expansion limit with an unconstrained, fresh SMT variable
of the specified return type.
The type-hypothesis for each such variable is added to the type-hypothesis list, $T$,
and the function call instance that this variable replaces is added to a list of
function calls instances, $F$.
As described in Section~\ref{sec:arch.sound}, \smtlink{} produces a clause for
ACL2 to check to verify that each function call in $F$ returns a value of
the user-claimed type.
Replacing the function's return value with an unconstrained variable is
a simple form of generalization.

\begin{Program}[t]
  \caption{:expand hint}\label{prog:expand-hint}
\begin{lstlisting}[language=Lisp]
:hints(("Goal"
        :clause-processor
        (Smtlink '( ...
                   (:expand ((:functions ((fun1 type1p) (fun2 type2p) ...
                                          (funk typekp)))
                   (:expansion-level 1)))
                   ...))))
\end{lstlisting}
\end{Program}
The user controls function expansion by \smtlink{} with a \texttt{:expand} hint as shown in
Program~\ref{prog:expand-hint}.
Each function is specified with its return type, and the :expansion-level parameter specifies
the maximum depth to which any function will be expanded.
We write $G_F$ to denote the clause produced by expanding the function calls in $G_T$.

\smtlink also supports translating function calls in ACL2 into uninterpreted function instances.
For example,\smallskip\\
\rule{2em}{0ex}\texttt{(:uninterpreted-functions ((expt rationalp integerp rationalp)))}\smallskip\\
says that the function {\texttt{expt}} should be treated as an unintepreted function whose first
argument satisfies \texttt{rationalp}, whose second argument satisfies \texttt{integerp}, and
whose return value satisfies \texttt{rationalp}.
\smtlink{} records each uninterpreted function declaration in a list, \textit{U}, and each call in $F$.

The mechanisms for function expansion and uninterpreted functions are similar.
In particular, the replacement of a recursive function call with a fresh variable is a
weaker version of replacing it with an uninterpreted function.
On the other hand, we discovered that Z3 does not combine its theories of non-linear arithmetic
and uninterpreted functions:  if a formula includes an uninterpreted function, the non-linear arithmetic
solver is silently disabled.
Thus, in many cases, using fresh variables is preferred to using uninterpreted functions.
We are examining these trade-offs in examples of real proofs and expect to formulate a more
unified treatment of function expansion and uninterpreted functions in a future version of \smtlink{}.

\subsubsection{Adding Hypotheses}{\label{sec:arch.phase1.hyps}}
Often, the proof of a theorem may depend on results that have already been established in ACL2's logical world.
However, \smtlink{} only translates the current goal for the SMT solver.
In practice, this is critical: while it is tempting to give the SMT solver every constraint that might be relevant,
this would often cause the SMT solver to require more time or memory than is available for the proof.
A key feature of the integration of SMT solvers into a theorem prover is that the user can identify the
\emph{relevant} facts, and these can be included with \texttt{:hypothesize} hints as illustrated in
Program~\ref{prog:poly-of-expt}.
Of course, the user can include any term they like in these hints.
If the SMT solver discharges the clause, then each of the \texttt{:hypothesize} hints is returned as a subgoal.
If it corresponds to a previously proven theorem, then ACL2 will (usually) discharge it without any further
assistance.
We write $H$ to denote the set of all hypotheses
introduced by \texttt{:hypothesize} hints,
$\widehat{H}$ to denote the conjunction of the elements of $H$,
and $G_H = \widehat{H} \Rightarrow G_F = \neg \widehat{H} \vee G_F$
to denote the goal clause augmented with these hypotheses.

\subsubsection{Substitutions}{\label{sec:arch.phase1.subst}}
Proof goals may include terms that do not have a representation in the theories of the chosen SMT solver.
For example, the theorem in Program~\ref{prog:poly-of-expt} used the \texttt{expt} function that raises
its first argument to an arbitrary integer power and is not representable in Z3 which only supports
fixed-degree polynomials and rational functions.
Rather than abandoning the advantages offered by the SMT solver,
\smtlink{} allows the user to specify a replacement of offending sub-expressions by fresh variables
of the appropriate types.
All occurrences of the given sub-expression are replaced by the specified variable.
This is another example of generalization by replacing the return value of a function with a fresh variable.
It is quite common, in our experience, to combine these substitutions with \texttt{:hypothesize} hints
that constrain the values of these variables.
Furthermore, the type-hypothesis for each new variable is included in the type-hypotheses list, $T$,
and the substitutions are recorded in a list $S$.

These substitutions are the final step of the first phase of translation.
We write $G'$ to denote the result of this first phase, and refer to it as the ``expanded clause''.

\subsection{The second translation phase}\label{sec:arch.phase2}
Given an original goal, $G$, along with user provided hints,
the first translation phase produces an ``expanded goal'', $G'$;
a list of type-assertions, $T$; a list of functions to be treated as uninterpreted, $U$;
a list of function call instances, $F$; a list of additional hypotheses, $H$;
and a list of substitutions, $S$.
The second translation phase uses these to produce the variable declarations for the SMT solver
and the claim that the SMT solver is to discharge.
If the SMT solver shows that \texttt{(not (implies $H$ $G'$))} is unsatisfiable for valuations of the free variables
that satisfy the type-hypotheses, $T$, and the uninterpreted function definitions, $U$,
then \smtlink{} concludes that \texttt{(implies $H$ $G'$)} is a theorem.
Unlike the first phase, the results of the transformations performed in this second phase are not
returned to ACL2 to be verified.
Our design goal was to keep this part of the connection as simple as possible to avoid errors and
enable code inspection by cautious users.

\subsubsection{Types}\label{sec:arch.phase2.types}
For each free-variable, $x_i$, occurring in $G$ (and thus in $G'$) there should be a corresponding
type-assertion, $T_i$ that is a conjunct of $T$.
For each type assertion, $(\textit{typep}_i~\textit{var}_i)$,
\smtlink{} generates a corresponding variable declaration for the SMT solver.
For example,\\
\rule{2em}{0ex}\texttt(rationalp x)\\
translates to\\
\rule{2em}{0ex}\texttt{x = \_SMT\_.isReal("x")}\\
In our implementation, the Python interface to the SMT solver is in the form of an object, \texttt{\_SMT\_}.
For example, \texttt{\_SMT\_.isReal("x")} creates a real-valued, symbolic variable for the SMT solver that underlies \texttt{\_SMT\_}.
If a type-assertion is omitted, then an undeclared variable will
appear in the formula to be checked by the SMT solver, and the SMT solver will report an error and fail.

\begin{Program}[t]
  \caption{The irrationality of $\sqrt{2}$}\label{prog:sqrt-of-2-is-irrational}
\begin{lstlisting}[language=Lisp]
(defthm sqrt-of-2-is-irrational
  (implies (rationalp x) (not (equal (* x x) 2))))

\end{lstlisting}
\end{Program}

For soundness, if \smtlink{} maps the ACL2 type recognizer $\textit{typep}_i$ to 
the SMT sort $\textit{sort}_i$, then every value that satisfies
$\textit{typep}_i$ must be an element of $\textit{sort}_i$.
Note that $\textit{sort}_i$ may include other values as well, this simply strengthens
the claim $G'$ and may result in a failure to prove a valid goal, but this will not cause \smtlink{}
to prove an invalid goal.
\smtlink{} maps the ACL2 type recognizers \texttt{booleanp} to SMT booleans; the type correspondence is strict.
The type recognizers \texttt{integerp} and \texttt{rationalp} are both mapped to SMT reals.
We did this because most SMT solvers (e.g.\ Z3) provide decision procedures for real numbers,
whereas ACL2 provides rationals.
As noted above, this strengthens the claim.
For example, the theorem shown in Program~\ref{prog:sqrt-of-2-is-irrational} can be proven in ACL2~\cite{Gamboa97},
we cannot discharge it using \smtlink{}.
It will report the counter-example for \texttt{x} equal to the square-root of two, described as an
algebraic number.
\smtlink{} can also be used with ACL2r, in which case the mismatch between rationals and reals can
be avoided entirely.

Our choice to broaden \texttt{integerp} to SMT rationals (instead of SMT integers) was pragmatic.
Our initial implementation uses the Z3 SMT solver, and we make extensive use of its non-linear
arithmetic solver.
Z3 disables the non-linear solver when a formula includes integer-valued variables.
By mapping ACL2 integers to SMT reals, \smtlink{} strengthens the theorem.
We expect to add mechanisms to allow the user to control whether ACL2 integers map to SMT integers
or reals in a future version of \smtlink{}.

\subsubsection{Functions}\label{sec:arch.phase2.functions}

The nine functions supported are
  \texttt{binary-+}, \texttt{unary--}, \texttt{binary-*}, \texttt{unary-/}, \texttt{equal},
  \texttt{<},        \texttt{if},     \texttt{not},    and \texttt{lambda}
along with the constants \texttt{t}, \texttt{nil}, and arbitrary integer constants.
As in ACL2, integers in Python can be arbitrarily large;
thus, \smtlink{} translates them directly.
\smtlink{} translates ACL2 lambda expressions into Python lambda expressions.
The other eight functions are translated directly to their counterpart methods of the \texttt{\_SMT\_} object.
For example, the ACL2 function \texttt{binary-+} is mapped to \texttt{\_SMT\_.plus}.
\smtlink{} generates declarations for all uninterpreted functions, again using the \texttt{\_SMT\_} interface.

If $G'$ includes any functions that are not in the list of eight above or in $U$,
then \smtlink{} will not prove $G$ but instead will fail with an error message.
In particular, unexpanded occurrences of user-defined functions will create an error.
Furthermore, any type-recognizer such as \texttt{rationalp} in $G'$ will create an error --
\smtlink{} requires that all type-recognizer terms occur in contexts that it can recognize as
type-hypotheses; others generate errors.
Likewise, $G$ cannot include quantification operators such as \texttt{exists} or \texttt{forall}.
This ensures that all variables appearing in $G'$ are free which is essential for our approach
of using SMT sorts that are super-sets of their ACL2 equivalents.
For example, one cannot state a theorem that 2 has a rational square root and ``prove'' it using
\smtlink{} to find a real-valued \texttt{x} such that \texttt{x*x $=$ 2}.

In the SMT world, each operation (such as \texttt{+}) is defined for specific sorts for its arguments
and defined to to produce a (symbolic) value of a specific sort for its result.
Some of these operations (such as \texttt{+}) are overloaded to operate on multiple types.
If an operator is applied to arguments for which it is not defined, then the SMT solver fails,
and \smtlink{} fails to prove the goal.
For example, if the original goal, $G$, (and thus $G'$) includes a term of the form \texttt{(+ x b)}
where \texttt{x} is real and \texttt{b} is boolean, then the SMT solver will fail even though the
operation is defined in ACL2.
This interpretation of ACL2 operators is conservative: \smtlink{} will not discharge an invalid theorem
due to the type restrictions of operators in the SMT world.

\smtlink{} translates \texttt{(/ $m$)} in ACL2 to \texttt{\_SMT\_.reciprocal($m$)}, where the
SMT function divides the constant 1 by $m$.
If $m \neq 0$, the ACL2 and SMT operations are identical.
If $m = 0$, then the SMT version produces an unconstrained integer (if $m$ is an integer) or real (if $m$ is real).
The ACL2 operator is defined to return 0.  Because the SMT version allows the ACL2 semantics, the SMT version
is more general.  Thus, \smtlink{} proves a more general claim, and a proof of $G'$ implies a proof of $G$.
This relies on our restriction that $G$ cannot include quantification operators.

\subsubsection{Hypotheses and Substitutions}{\label{sec:arch.phase2.hyps}}
These are handled entirely in the transformation of the original goal, $G$, to the expanded goal, $G'$,
in the first translation phase and do not impact the second phase.

\subsection{Ensuring soundness}\label{sec:arch.sound}
Our design goal with \smtlink{} has been to trust ACL2,
the chosen SMT solver (Z3, in our current implementation), and as little other code as practical.
At the same time, our intended use for \smtlink{} is for the verification of AMS circuits
and other cyber-physical systems.
Because we are developing verification techniques as we go, we want \smtlink{} to provide a flexible
framework for prototyping new ideas.
Our solution is to put most of the functionality and complexity of \smtlink{} into the first translation phase.
If the SMT solver discharges the translated clause, then \smtlink{} generates a set of return
clauses to check the correctness of this translation.
The second phase is trusted; this code is both small and simple.


Our basic approach is simple:
let $A$ denote the additional assumptions that were added to the goal by type assertions for variables
and function return values, hypotheses, and substitutions.
Let $G_{\mathit SMT}$ denote the clause that is tested by the SMT solver.
If the SMT solver proves $G_{\mathit SMT}$, then \smtlink{} returns the clauses
\begin{equation}\label{eq:defQ}\begin{array}{rcl}
  Q_1 &=& (G' \wedge A) \Rightarrow G\\
  Q_2 &=& A \vee G
\end{array}\end{equation}
for proof by ACL2.
We are trusting the translation of $G'$ to $G_{\mathit SMT}$ and the SMT solver itself,
Modulo that trust, the truth of $G_{\mathit SMT}$ implies the truth of $G'$;
in which case $Q_1$ is equivalent to $A \Rightarrow G$.
Accordingly, when ACL2 proves $Q_1$ and $Q_2$, $G$ is established as a theorem.

We make two observations before describing how each step of the translation process contributes to $A$.
First, the correctness of this argument does not depend on the
choice of $A$.  Of course, deriving the intended $A$ is important to ensure that $Q_1$ and $Q_2$ can actually
be proven.
Second, $A$ is the conjunction of the various assumptions that were added by \smtlink{}.
\smtlink{} expresses $Q_2$ as a separate subgoal for each conjunct of $A$.

\subsubsection{Types}\label{sec:arch.sound.types}
Each type-hypothesis identified by \smtlink{} is included in $A$.
Let $T_i = (\textit{typep}_i~\textit{var}_i)$ be such a type-hypothesis.
When proving $Q_2$, ACL2 verifies $T_i \vee G$
which means that for all values of $\textit{var}_i$ that do not satisfy $\textit{typep}_i$,
$G$ trivially holds.
By the trust that \smtlink{} declares $\textit{var}_i$ to be of an SMT sort that includes all values that satisfy
$\textit{typep}_i$, the translation is valid.

\subsubsection{Functions}\label{sec:arch.sound.functions}
When a function call is expanded in the first translation phase, the equivalence is checked by ACL2 when it
verifies $Q_1$.
We are trusting the translation of ACL2 lambda-expressions to their Python equivalents in phase 2.
When a function call is replaced by a variable, ACL2 must check that the user-claimed type for the return value
of the function is valid.
This is done by generating a clause for each function call in $F$.
Let $f$ be such a function call (i.e.\ an ACL2 term), and let $\textit{type}_f$ be the user-claimed type for the
return value of $f$.
\smtlink{} includes a conjunct of the form
\begin{equation}\label{eq:return-type}
    (\texttt{or} ~ (\textit{type}_f ~ f) ~ G)
\end{equation}
in $Q_2$.
A technical detail is that $f$ may include variables that are bound by lambda expression arising from other
function expansions; such variables are free in the clause depicted in Equation~\ref{eq:return-type} as generated by \smtlink{}.
This means that these variables are less constrained in the check performed by ACL2 than they are in $G'$ or $\gsmt$.
Because ACL2 has proven the more general case, we can safely conclude the more restricted version as well.

\subsubsection{Added Hypotheses}{\label{sec:arch.sound.hyps}}
Each hypothesis, $H_i$, added by the user, is included in $A$.
The clause $(H_i \vee G)$ is verified by ACL2;
therefore, it is safe to add $H_i$ as a hypothesis for $G'$ (and thus for $\gsmt$).

\subsubsection{Substitutions}{\label{sec:arch.sound.subst}}
\smtlink{} records the  user-defined substitutions in the list $S$.
When generating $Q_1$ and $Q_2$, \smtlink{} uses lambda expressions to bind the variables declared in
substitution hints to their corresponding expressions --
this is similar to the way that function expansions are handled.
Furthermore, the user-claimed types of these expressions are included in $T$, and \smtlink{} generates
clauses for ACL2 to check these claims in the same manner as checking the types of values returned by
function calls.

\subsubsection{The Python Interface}{\label{sec:arch.sound.python}}
\smtlink relies on software packages that are outside the ACL2 world, namely the Python interpreter and
an SMT solver (Z3 for the purposes of this paper).
This creates the potential unsoundness that these external components can be modified without detection.
Our implementation of \smtlink{} takes several measures to prevent the most likely causes of unsoundness.
First, \smtlink{} has a default configuration that is encoded in \texttt{config.lisp}.
There is a script for creating \texttt{config.lisp}; once run, the configuration includes full path
names to the Python interpreter and sets the path variable for searching for Python classes.
Likewise, the Python code to define the class for the interface object, \texttt{\_SMT\_} described
in Section~\ref{sec:arch.phase2} is provided as the string returned by the function ACL22SMT.
The file ACL22SMT.lisp is generated from a Python source file that is specific for the intended SMT solver.
The consequence of this approach is that the paths to the Python interpreter and the SMT solver (and
therefore the choice of the SMT solver), along with the Python class definition for the interface
between \smtlink{} and the SMT solver are all baked into the certified ACL2 code for \smtlink.
We believe that this should make \smtlink{} quite robust to unintentional changes of the computing
environment.  Of course, a nefarious user could replace the executable image for the Python interpreter,
or the dynamic library for the SMT solver, but these are in ``system'' directories (under \texttt{/usr/bin}
in our installation) rather than user directories; so such changes are unlikely to be accidental.
Such changes \emph{are} likely to occur as a consequence of regular software updates.  We are
considering adding checksum information to our \texttt{config.lisp} to ensure that such changes are
detected and reported.  We would like to devise an SMT-solver independent way of recording such
checksums.

\subsubsection{Remarks}{\label{sec:arch.sound.remarks}}
In the current implementation of \smtlink{}, the construction of the goals $Q_1$ and $Q_2$ is done
within the trusted code of the clause processor.
Although the arguments for the correctness of these constructions are straightforward,
the fact that this is unverified code does present a risk of errors.
As we have learned more about ACL2, we now see that an alternative would be to restructure
\smtlink{} to provide a function that returns $G'$ and the lists $T$, $F$, $U$, $H$, and $S$ described above.
From these, a local theorem, that $G'$ holds would be proven using a trusted clause processor corresponding
to phase 2 of the current \smtlink{}.
Additional local theorems would be proven by ACL2 to prove $Q_1$ and each clause of $Q_2$ from Equation~\ref{eq:defQ}.
Then, the main theorem, $G$ would be proven by ACL2 using these local theorems.
This should be a relatively straightforward restructuring \smtlink{} that would isolate the small amount
of trusted code.
We plan to do this in the near future.

Even greater confidence could be achieved by adopting the ``skeptical'' approach advocated
by Harrison~\cite{Harrison98}, for example by using
proof reconstruction~\cite{Blanchette13,Fontaine06,Mclaughlin06,Armand11,Merz12} or proof certificates~\cite{Besson06}.
We see such efforts as complementary to the approach that we have taken with \smtlink{}.
We are using \smtlink{} to develop proof methods for domains where formal methods have had little prior use.
As described in Section~\ref{sec:custom}, the relatively lightweight interfaces in \smtlink{} facilitate
such experimentation.
We gain this flexibility at the risk that an error in critical parts of our code (or in the
SMT solver itself) could lead to a ``proof'' of a non-theorem.
We believe that this risk is small compared with other risks that are inherent in the verification of
physical artifacts: most notably, ``Does the model of the physical system actually capture all possible
behaviours?''  Being able to prototype and develop proofs quickly lets us explore the consequences of
the models more thoroughly than would be possible with a less flexible approach.
Thus, we regard the slight risk of an error as being justified by the opportunity to verify designs
that are otherwise outside the reach of formal tools.
We see this as complementary to work on proof reconstruction and proof certificates.
If we demonstrate the kinds of proofs that are useful in practice, that should illuminate where
proof reconstruction and certificates would offer the greatest increase in confidence in critical designs.

\section{Customizing \smtlink}\label{sec:custom}
The design choices described in Section~\ref{sec:arch.sound.python} protect the user from unintentional changes to the external components
of \smtlink.
What if such changes are desired?
To facilitate such experimentation, we provide a second version of \smtlink, \cnfsmtlinkc, where the
user can easily change the configuration of external components.
Using \cnfsmtlinkc{} requires a different trust-tag than that for the standard configuration,
\cnfsmtlink{}.
Thus, it is easy to track theorems whose proofs descend from a custom configuration of the clause processor.
The remainder of this section describes one such custom configuration to illustrate how these features
facilitate experimentation.

Our largest use of \smtlink{} to date has been the proof of global convergence for a digital phase-locked
loop (The code can be found at~\cite{DPLLproof} and see~\cite{Peng15a} for more details.).  The original proof was a 13 page long latex document, with
lots of tedius algebra.  Using the standard configuration of \smtlink{}, we completed the same proof
using ACL2.  The proof is about 1700 lines of ACL2 code.  While \smtlink{} made the proof possible,
it didn't make it as easy as we had hoped.  A key complication is that the phase-locked loop (PLL)
model uses recurrence functions whose solutions make extensive use of ACL2's \texttt{expt} function.
As described earlier, \smtlink{} can handle these, but each occurrence requires \texttt{:let} and
\texttt{:hypothesize} hints.  Furthermore, function expansion renames variables; so, the proofs
involved many lemmas whose sole purpose was to explicitly expand functions and rewrite terms so
as to make the calls to \texttt{expt} visible in the theorem statement and thus amenable to these hints.

Our solution was to define a new Python class for the \texttt{\_SMT\_} interface object.
This class is called \texttt{RewriteExpt}, and it extends the default \texttt{ACL2\_to\_Z3} that was
complied into \texttt{ACL22SMT.lisp} as described above.
To use this extension, \texttt{expt} is declared to be an uninterpreted function.
\texttt{RewriteExpt} overrides the \texttt{\_SMT\_.prove} method to add a pre-processing step finds
instances of \texttt{expt} in the claim.
For each instance, the code checks to see if the hypotheses of the theorem imply the guard for \texttt{expt}:
the base must be non-zero, or the exponent must be non-negative.  If the guard can't be proven, an error is
reported and the proof fails.  Otherwise, \texttt{RewriteExpt} applies a small number of simple proof
rules about \texttt{expt}.
If the antecedent of one of these rules is satisfied, then the consequent is added as a new hypothesis.
Table~\ref{tab:expt-rules} shows some examples of these rules.

\begin{table}\caption{Rules for \texttt{expt}}\label{tab:expt-rules}
\begin{center}\tt\begin{tabular}{rll}\hline
    1. & (expt x 0) $\rightarrow$ 1\\
    2. & (expt 0 n) $\rightarrow$ 0, & if $\texttt{n} > 0$\\
    3. & (expt x (+ n1 n2)) $\rightarrow$ (* (expt x n1) (expt x n2))\\
    4. & (expt x (* c n)) $\rightarrow$ (* (expt x n) (expt x n) \textrm{\ldots} (expt x n))\\
    5. & (< (expt x m) (expt x n)), \textrm{if $1 < \texttt{x}$ and $\texttt{m} < \texttt{n}$}\\
    6. & \ldots\\ \hline
\end{tabular}\end{center}
Notes: All rules have a precondition of that either the base is non-zero or the exponent is positive;
  furthermore, new instances of \texttt{expt} are only generated if they can be shown to satisfy the
  same condition.
  For rule 4, the right-hand side of $\rightarrow$ is the multiplication of \texttt{c} copies of \texttt{(expt x n)}. 
  Rule 4 is only applied if \texttt{c} is small and positive.
\end{table}

Preliminary experiments with this customized clause processor have been very promising.
For example, one theorem in the PLL proof that required 19 supporting lemmas for
a total of 334 lines of ACL2 code was replaced by a single theorem stated in 13 lines of ACL2 code.
The proofs with the customized clause processor are much shorter, much simpler, and much easier to understand.

We are in the process of writing a new proof for the PLL based on the customized clause processor.
We see many directions that we could pursue to extend this approach after revising the PLL proof.
First, the customized clause processor uses a set of proof-rules that are hard-coded into
\texttt{RewriteExpt.py}.  These correspond to runes for existing ACL2 theorems about \texttt{expt}.
We expect that we could forward such runes from ACL2 to the SMT interface and write a simple, generic
inference engine in Python.  The advantage of performing the inference with the SMT solver is that it
can discharge pre-conditions for runes that ACL2 does not resolve with its waterfall.
On the other hand, the ACL2 framework is much more general than what can be described in the theories
of an SMT solver; so we see the two as complementary.
We also note that once our inference engine has discovered a useful hypothesis, it also has the
justification.  Thus, we could return these to ACL2 and use them to generate the
\texttt{:let} and \texttt{:hypothesize} needed to discharge the goal with the standard configuration
of \smtlink.  If this approach were implemented, then our customized processor would be an elaborate
computed hint, but the goal would be discharged with \cnfsmtlink, and no additional trust would be
required.

\section{Related work}\label{sec:relwk}
There has been extensive work in the past decade on integrating SAT and SMT solvers into theorem provers.
Srinivasan \cite{srinivasan2007} integrated the Yices~\cite{Dutertre2014} SMT solver into ACL2 for verifying
bit-level pipelined machines.
They also use the mechanism of a trusted clause processor with a translation process quite similar to ours.
They appear to have mostly used the bit-vector arithmetic and SAT solving capabilities of Yices.
While they also produce an expanded formula that is then translated to SMT-LIB~\cite{Barrett10},
they don't describe using ACL2 to check this translation as we have done.
Prior to that, in~\cite{Manolios2006}, they integrated a decision procedure called UCLID~\cite{Lahiri2004} into ACL2 to solve a similar problem.

Works on integrating SMT solvers or techniques into other theorem provers include \cite{Mclaughlin06,Fontaine06,Besson06,Armand11,Merz12,Blanchette13,Deharbe14}. Many of these papers have followed
Harrison and Th\'{e}ry's ``skeptical'' approach~\cite{Harrison98} and focused on methods for verifying SMT
results within the theorem prover using proof reconstruction, certificates, and
similar methods. Several of the papers showed how their methods could be used
for the verification of concurrent algorithms such as clock synchronization \cite{Fontaine06},
and the Bakery and Memoir algorithms \cite{Merz12}. While \cite{Fontaine06} used the CVC-Lite \cite{Barrett04}
SMT solver to verify properties of simple quadratic inequalites, the use of SMT
in theorem provers has generally made light use of the arithmetic capability
of such solvers. In fact \cite{Blanchette13} (Isabelle/Sledgehammer with Z3) reported better results for SMT for several sets of
benchmarks when the arithmetic theory solvers were disabled!

The work that resembles our approach is \cite{Deharbe14};
they present a translation of Event-B sequents from Rodin \cite{Abrial06} to the SMT-LIB format \cite{Barrett10}.
Like our work, \cite{Deharbe14} verifies a claim by using a SMT solver to show that its negation is unsatisfiable.
They address issues of types and functions.
They perform extensive rewriting using Event-B sequents, and then have simple translations of the rewritten form into SMT-LIB.
While noting that proof reconstruction is possible in principle, they do not appear to implement such measures.
The main focus of \cite{Deharbe14} is supporting the set-theoretic constructs of Event-B.
In contrast, our work shows how the procedures for non-linear arithmetic of a modern SMT solver can
be used when reasoning about analog and mixed-signal circuits.

Our work demonstrates the value of theorem proving combined with SMT
solvers for verifying properties that are characterized by functions on real numbers
and vector fields. Accordingly, the linear and non-linear arithmetic theory
solvers have a central role. As our concern is bringing these techniques to new
problem domains, we deliberately take a pragmatic approach to integration and
trust both the theorem prover and the SMT solver.

Prior work on using theorem proving methods to reason about dynamical
systems includes \cite{Immler14} which uses the Isabelle theorem prover to verify bounds on
solutions to simple ODEs from a single initial condition. Harutunian \cite{Harutunian07} presented a very
general framework for reasoning about hybrid systems using ACL2 and demonstrated
the approach with some very simple examples. Here we demonstrate that
by discharging arithmetic proof obligations using a SMT solver, it is practical
to reason about much realistic designs.

\section{Conclusion and future work}\label{sec:concl}
This paper presented \smtlink{}, a clause-processor that we have used to integrate the Z3 SMT solver into ACL2.
Reasoning about systems of polynomial and rational function equalities
and inequalities can be greatly simplified by using Z3's non-linear arithmetic capabilities.
ACL2 complements Z3 by providing a versatile induction capability along with
a mature environment for proof development and structuring.
\smtlink{} offers two configurations: the default, standard configuration where the interface code and
the pathways to the external tools (Python and Z3) are fixed when book is certified; and a customizable
interface that allows the user to experiment with extending these capabilities.

Section~\ref{sec:arch} described our software architecture, issues that arose when integrating an SMT
solver into ACL2, and our solutions to these issues.  A key aspect of the design is a two-phase translation
process for converting ACL2 clauses into formulas that can be discharged by the SMT solver.
The first phase translates a fairly expressive subset of ACL2 into a simple subset consisting of nine
built-in functions.
This first phase includes methods for handling types, function expansion, uninterpreted functions,
and sub-expression replacement; all of these can be understood as various versions of generalizing
the original clause to produce a stronger clause that is suitable for discharging with an SMT solver.
Most of the complexity of the translation process is in the first phase.
Because ACL2 verifies that the clause produced by this first phase implies the original, this first
phase greatly improves the usability of the clause processor while raising minimal concerns about soundness.
The second phase transliterates the nine remaining functions to equivalents in a Python API -- this
is the code that is most critical for sondness.

Section~\ref{sec:custom} showed how the customizable interface can be used to automate tedious aspects of
a moderately large (1700 line) proof that we performed with the original version of the clause processor.
By adding a few simple rules for transforming expressions involving the ACL2 \texttt{expt} functions into
the SMT interface, we showed that we could dramatically reduce the length and complexity of some of the
proofs.  We believe that this demonstrates the value of \smtlink{} as an experimental platform.
Once a proposed functionality is shown to have sufficient value, then a more rigorous version could
be implemented.  The fast prototyping that is enabled by \smtlink{} can help guide this process by
avoiding investing large amounts of effort on some approach that ultimately provides small improvements
to the proof development process.

Prior work on integrating SMT solvers into theorem provers has focused on using the non-numerical
decision procedures of an SMT solver.
Our work focuses on the value of bringing an SMT solver into a theorem prover for
reasoning about systems where a digital controller interacts with a continuous,
analog, physical system.
The analysis of such systems often involves long, tedious, and error-prone derivations
that primarily use linear algebra and polynomials.
These are domains where SMT solvers combined with induction and proof structuring have great promise.

\subsection{Future work}\label{sec:concl.future}

\smtlink{} returns clauses to ACL2 to check the translation of the original goal to a small subset of ACL2.
As noted in Section~\ref{sec:arch.sound.remarks}, a moderate restructuring of this code could allow most
of this work to be done within ACL2 and reduce the amount of code that must be trusted in \smtlink{}.
We believe that this could be done with minimal impact on the flexibility of \smtlink{} for experimenting
with SMT solvers and their applications.

We have used ACL2 with \smtlink{} to prove the most challenging part of a global convergence argument for
a digital Phase-Locked Loop (PLL) using \smtlink.  Global convergence is a response property, and
we can show that the PLL makes progress through four distinct phases.  We used ACL2 with \smtlink{} to
verify the phase for which a hand-written proof was the most complicated.  We would like to write proofs
in ACL2 for the other three phases and use ACL2 to prove that those results are sufficient to prove
correct convergence from any initial condition.  This will involve constructing Skolem functions to
compose the individual pieces of the proof and should demonstrate the strength of using ACL2 to prove
properties that cannot be expressed in the logic of the SMT solver.

We would like to add a bounded model checking capability to SMT link.
For example, in the PLL proof, there is a tedious proof at the boundary of two of the phases.
Z3 provides an ``easy'' proof by showing that within eight steps of the recurrence, the transition between phases is complete
and correct.  We would lke to integrate this capability into \smtlink{} and thus into ACL2.

The current implementation of \smtlink{} provides very restricted support for recursive functions.
This is because most recursive functions are non-numerical and/or use ``fixing'' functions or recognizers
to ensure termination when called with bogus arguments.
While this has not been problematic for our PLL proof, we would like to generalize the handling of types
by \smtlink{} to allow a wider range of applications.

Many of the type-checking steps performed by \smtlink{} reconstruct facts that are already present in
ACL2s \texttt{type-alist}.  We would like to see if this information could be used by \smtlink{} and
thus spare the user of many of the type declaration that \smtlink{} now requires.

Presently, \smtlink{} prints counter-examples from the SMT solver to the ACL2 comment window.
We would like to make them available to the user within the ACL2 environment.
This could be similar to the \texttt{env\$} argument used in \texttt{Satlink}.
New issues arise in the SMT case because SMT formulas don't have a single, syntactical form like CNF for SAT.
Furthermore, if the counter-example included irrational numbers, then it cannot be represented in ACL2 --
although this should be addressed in ACL2(r).

\subsubsection*{Acknowledgments}
We would like to thank the many members who have patiently answered our many questions while developing \smtlink{},
especially Matt Kaufmann and David Rager. We are also thankful to the anonymous reviewers for the insightful and inspiring feedback.

\nocite{*}
\bibliographystyle{eptcs}
\bibliography{ref}
\end{document}